\documentclass[%
aps,
prl,
twocolumn,
superscriptaddress,
amsmath,
amssymb,
citeautoscript
]{revtex4-1}
\usepackage{graphicx}
\usepackage{dcolumn}
\usepackage{bm}
\usepackage{hyperref}

\usepackage{xcolor}
\usepackage{soul}
\usepackage[normalem]{ulem}

\newcommand{\moviesolid}{S1}
\newcommand{\moviefluid}{S2}
\newcommand{\moviecomplete}{S3}
\newcommand{\moviesinglecomplete}{S4}
\newcommand{\moviesinglescutoid}{S5}

\newcommand{\siheight}{S1}
\newcommand{\siapicobasal}{S2}
\newcommand{\silinear}{S3}

\newcommand{\enmdim}{Appendix A}
\newcommand{\enmforces}{Appendix B}
\newcommand{\enmfluid}{Appendix C}
\newcommand{\enmsingle}{Appendix D}

\begin{document}

\preprint{APS/123-QED}

\title{Full Three-dimensional Cell Geometry Matters for Epithelial Fluidization}

\author{Toshikaze Chiba}
\affiliation{Jožef Stefan Institute, Jamova 39, SI-1000 Ljubljana, Slovenia.}

\author{Jan Rozman}
\affiliation{Jožef Stefan Institute, Jamova 39, SI-1000 Ljubljana, Slovenia.}
\affiliation{University of Ljubljana, Faculty of Mathematics and Physics, Jadranska 19, SI-1000 Ljubljana, Slovenia.}

\author{Matej Krajnc}
\email{matej.krajnc@ijs.si}
\affiliation{Jožef Stefan Institute, Jamova 39, SI-1000 Ljubljana, Slovenia.}

\date{\today}
\begin{abstract}
Epithelial dynamics are often studied using two-dimensional models that omit cell height. Here, we investigate how this third dimension influences tissue fluidization driven by cell self-propulsion. We identify three effects absent from conventional 2D models. First, cell height controls fluidization through a reentrant transition: squamous and columnar tissues are more resistant to fluidization than cuboidal ones. Second, neighbor exchanges through transient scutoid states substantially enhance cell motility, compared to classic T1 transitions. Third, self-propelling cells develop pronounced apicobasal shape asymmetry, with the experimentally less accessible basal surface providing a more accurate indicator of cells' target shape index and motility.
\end{abstract}

\maketitle

{\textit{Introduction}}---The ability of epithelial cells to rearrange and exchange neighbors is central to tissue remodeling and collective behavior. Depending on their mechanical properties and activity, epithelial monolayers can exhibit either solid-like states or fluid-like behavior, characterized by frequent cellular rearrangements. These phenomena are commonly investigated theoretically using simplified models in which the monolayer is represented in only two dimensions. Such approaches include cellular Potts models~\cite{graner1992simulation,glazier1993simulation,hirashima2017cellular}, multiphase-field models~\cite{nonomura:2012,najem:2016,mueller:2019}, and particle-based models~\cite{szabo2006phase,sepulveda2013collective,garcia2015physics}.

A particularly common approach is the 2D vertex model~\cite{Farhadifar07,fletcher2014vertex,alt2017vertex} which parametrizes cell shapes by the positions of their vertices. Its canonical formulation, which encodes area-and-perimeter elasticity (APE)~\cite{Farhadifar07}, features a rigidity transition, whereby the mechanical response of the tissue changes from rigid to floppy and the energy barriers for T1 transitions vanish as the cells' target shape index (perimeter over root area) $p_0$ increases past a threshold $p_0^*$~\cite{Farhadifar07,Bi15}. The value of $p_0^*$ was initially reported as $3.81$~\cite{Bi15}, while subsequent studies found somewhat higher values~\cite{merkel2019minimal,wang2020anisotropy,tong2022linear}.

Various generalizations of the 2D vertex model incorporate active forces, including self-propulsion~\cite{Bi16,barton2017active,koride2018epithelial,lin2018dynamic,amiri2023random}, nematic activity~\cite{comelles2021epithelial,lin2023structure,rozman2024cell}, and junctional fluctuations~\cite{curran2017myosin,Krajnc20,Yamamoto22}. For example, in models of self-propelled tissues, cell motility is driven by active crawling along a substrate~{\cite{Alert2020-oj}}, with the direction of self-propulsion typically assumed to undergo rotational diffusion~\cite{Bi16}. The transition from a solid-like tissue of caged cells to a fluid-like tissue with motile cells is then controlled by the strength of the active force, the target shape index $p_0$, and the rotational diffusion coefficient. Crucially, any active force can fluidize a tissue with $p_0$ in the floppy regime, but increasingly stronger activity is needed for fluidization as $p_0$ moves further into the rigid regime. 

While tissue models are often 2D, real cells are of course not. They have distinct apical, basal, and lateral surfaces with different properties. For example, cortical actomyosin can polarize to form contractile structures that are typically enriched at the apical surface~{\cite{Campas2024-vf}}, whereas traction forces that drive cell migration on substrates are exerted basally~{\cite{Alert2020-oj}}. Moreover, apical and basal cell-neighbor relationships need not coincide, allowing for scutoid shapes~\cite{Rupprecht2017-lk,GomezGalvez18}.

This additional cell-height dimension can profoundly affect the \emph{passive} mechanical properties of epithelial tissues. A minimal three-dimensional extension of the 2D area-and-perimeter elasticity model that accounts for the apicobasally polarized architecture of epithelial cells eliminates the rigidity transition, leaving the tissue solid-like~\cite{rozman2024basolateral}. By contrast, adapting a model that encodes volume-and-surface-area elasticity~\cite{merkel2018geometrically} to an epithelial monolayer gives rise to a cell-height-dependent reentrant rigidity transition~\cite{Sarkar2026-ql}.

In this Letter, we investigate how 3D effects influence the behavior of \emph{active} tissues using a 3D vertex model with volume-and-surface-area elasticity and basal cell self-propulsion. We identify three processes that rely on the cell-height dimension and are not directly captured by the 2D model: (i)~cell height controls a reentrant solid-fluid-solid transition, such that cuboidal tissues are easier to fluidize than either their squamous or columnar counterparts; (ii)~3D cell rearrangements through the scutoid pathway can significantly increase cell motility compared to the standard T1 transition; and (iii)~fluid-like tissues can develop pronounced apicobasal asymmetry, in contrast to the symmetric cell shapes implicitly assumed in 2D models. Although experiments typically image the apical surface, we find that basal cell geometry provides a more direct, cell-height-independent readout of the underlying physical parameters.

\begin{figure*}[t]
    \centering
    \includegraphics[width=\textwidth]{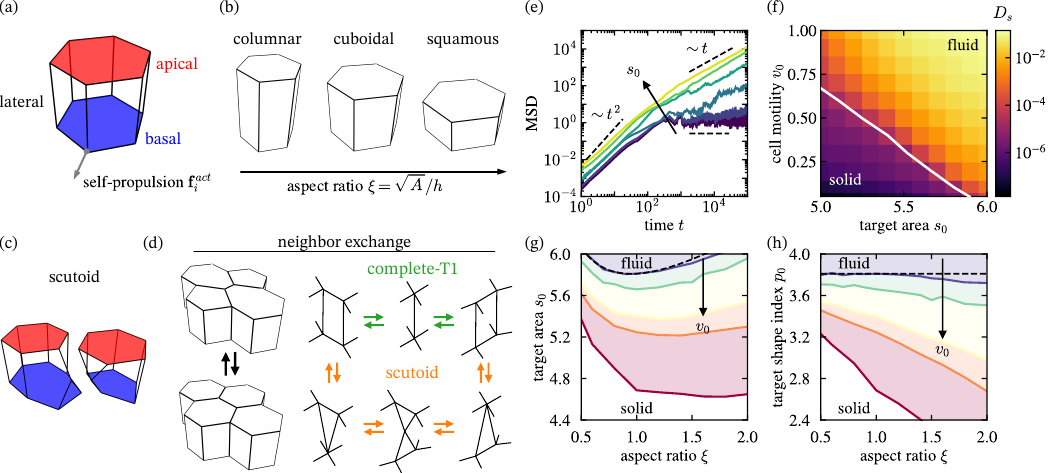}
    \caption{\label{fig:schematics}
    (a) Cell in the 3D vertex model, with apical (red) and basal (blue) side colored and the self-propulsion force illustrated. (b) Cell aspect ratio $\xi$ controls the columnar-to-squamous transition. (c) Schematic of the scutoid shape: The cell has a different number of apical and basal neighbors. (d) Schematic of two possible neighbor exchanges routes. In a complete-T1 case (green), cells simulatenously exchange neighbors on the basal and apical side, corresponding to a T1 in the 2D model. In the scutoid case (orange), a cell first exchanges neighbors on one side, forming a scutoid. This can then progresses in a zipper-like fashion to the opposite side, concluding the neighbor exchange. (e) MSD curve for $s_0 \in \{5.0, 5.2, 5.4, 5.6, 5.8, 6.0\}$ at $v_0 = 0.5$ and $\xi = 1$. 
    (f) Heatmap of the cells' diffusion coefficient at $\xi=1$ as a function of $s_0$ and $v_0$. White line separates solid-like and fluid-like regions. 
    (g),(h) Solid-fluid boundary at $v_0\in\{0.1, 0.2, 0.4, 0.5,1.0\}$ as a function of cell aspect ratio $\xi$ and either the target surface area $s_0$ (g) or the effective target shape index $p_0$ [Eq.~\eqref{eq:p0}] (h). Dashed line on both panels shows the rigidity transition boundary in the passive  model ($v_0=0$)~\cite{Sarkar2026-ql}.
    }
\end{figure*}

\begin{figure*}[t]
    \centering
    \includegraphics[width=\linewidth]{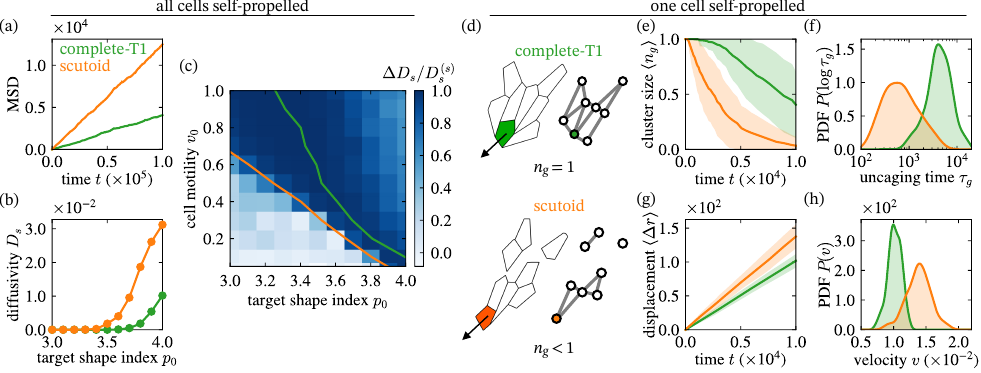}
    \caption{\label{fig:scutoid}
    (a) $\mathrm{MSD}$ for the complete-T1 and scutoid models at $s_0=6$, $v_0=0.5$, and $\xi=1$. Legend applies to all panels. (b) Diffusion coefficient as a function of $p_0$ [Eq.~\eqref{eq:p0}] for both models. (c) Heatmap of the normalized difference in diffusivity between the complete-T1 and scutoid models at $\xi=1$: The scutoid model is consistently more motile. Green and orange lines show the fluidization threshold for the complete-T1  and scutoid model, respectively. (d)-(h) Single self-propelled cell model. (d) Example of a self-propelling cell (colored) at $t=900$ in both models. Also drawn are its initial neighbors and corresponding graph of connected cells. The complete-T1 active cell has remained in a connected region with all initial neighbors ($n_g=1$); the scutoid active cell has not ($n_g<1$). (e)-(h) Comparison of the complete-T1 and scutoid model: (e) average fractional cluster size $n_g$ as a function of time; (f) distribution of uncaging time in different simulations; (g) displacement from the starting position as a function of time; (h) self-propelled cell velocity distribution between different simulations. Panels (d)-(h) use $s_0=6.2$, $v_0=0.5$, and $\xi=1$. Data on panels (e)-(h) is averaged over 100 simulations with different randomly selected self-propulsion directions; shaded regions on panels (e),(g) show standard deviation. See \enmsingle\ for details of how these values were calculated.}
\end{figure*}


{\textit{3D monolayer model}}---We consider a three-dimensional confluent monolayer of self-propelled cells described by a vertex model. Each cell is represented as a polyhedron whose geometry is defined by vertex positions. One face of each cell forms its apical surface, another its basal surface, while the remaining faces constitute the lateral cell-cell interfaces [Fig.~\ref{fig:schematics}(a)]. For simplicity, apical and basal vertices are confined to two parallel planes separated by a fixed height [Fig.~\ref{fig:schematics}(b)]. The model allows for scutoids~\cite{GomezGalvez18, Rupprecht2017-lk}, so that the apical and basal polygons can have a different number of sides~[Fig.~\ref{fig:schematics}(c)].

We assume that tissue mechanics are governed by volume-and-surface-area elasticity (VAE). Taking $V_0^{1/3}$ and $K_SV_0$ as the units of length and force, respectively, where $V_0$ is the target cell volume and $K_S$ is the surface-area elasticity modulus, the dimensionless energy reads
\begin{equation}
    w=\sum_k\left[k_v\left(v_k-1\right)^2+\left(s_k-s_0\right)^2\right],
    \label{eq:energy}
\end{equation}
where $v_k$ and $s_k$ denote the dimensionless volume and surface area of cell $k$, respectively, $s_0$ is the dimensionless target surface area, and $k_v$ is the dimensionless volume elasticity modulus~\cite{merkel2018geometrically,Sarkar2026-ql}. Details of the nondimensionalization are provided in \enmdim.

Vertices follow overdamped dynamics described by
\begin{equation}
    \dot{\mathbf r}_i=\mathbf f_i^{cons}+\mathbf f_i^{act}+\mathbf f_i^{aux},
\label{eq:vertex_eom}
\end{equation}
where $\mathbf{r}_i$ is the dimensionless position of vertex $i$, the overdot denotes the time derivative, and unit of time is defined by the interplay between friction (i.e., inverse mobility $\mu^{-1}$) and mechanical tension: $\tau_0=(K_SV_0^{2/3}\mu)^{-1}$. On the right-hand-side, $\mathbf f_i^{cons}=-\nabla_{\mathbf{r}_i}w$ and $\mathbf f_i^{act}$ are the dimensionless conservative force and active self-propulsion force of vertex $i$, respectively; $\mathbf f_i^{aux}$ is an auxiliary force that prevents the apical and basal surfaces from collapsing (\enmforces).

To model active self-propulsion in analogy with the 2D approach~\cite{Bi16}, each cell $k$ is assigned a polarity angle $\theta_k$, which undergoes angular diffusion with rotational diffusion coefficient $D_r=0.01$. The active force on the basal side $\textbf{f}_k^{act,cell}=v_0 \left[\cos(\theta_k),\sin(\theta_k)\right]$ is then translated into forces on basal vertices (\enmforces); $v_0$ sets the strength of the active force. 

Cells in the model exchange neighbors through edge-to-triangle (ET) and triangle-to-edge (TE) transitions, which together constitute a complete T1 transition via an intermediate scutoid configuration [Fig.~\ref{fig:schematics}(d)]~\cite{okuda2013reversible,sarkar2024graph}. An ET transition is triggered when an edge becomes shorter than the threshold $l_{\rm th}=0.01$. Similarly, a TE transition occurs when an apical or basal edge of a lateral triangle falls below the same threshold. In both cases, the newly formed edge is assigned a length of $0.011$.

The height $h$ separating the apical and basal planes is set by the cell aspect ratio $\xi$. Assuming an apicobasally symmetric cell shape with volume $1$, its apical (and basal) area is $A=1/h$, its aspect ratio is $\xi=\sqrt{A}/h=h^{-3/2}$, and therefore $h=\xi^{-2/3}$ [Fig.~\ref{fig:schematics}(b)]. The $x$- and $y$-dimensions of the simulation box, $L_x=L_y=L$, are then chosen to fit the total target volume of all cells between the planes, $L=\xi^{1/3}\sqrt{N_c}$, where $N_c=1024$ is the cell number. Simulations have periodic boundary conditions along $x$ and $y$. Throughout the simulation, we use $k_v=10$, so that cells are treated as nearly incompressible. The equations of motion are solved using an Euler scheme with time step $\Delta t = 0.01$.

{\textit{Height-mediated reentrant fluidization}}---We now explore how basal self-propulsion fluidizes a 3D monolayer. We first focus on the case of $\xi=1$, $s_0=5$, corresponding to a cuboidal tissue deep in the rigid regime~\cite{Sarkar2026-ql}, and set $v_0=0.5$. For these parameters, cells remain caged by their neighbors and the tissue does not fluidize (Movie~\moviesolid). To quantify this, we measure the cells'  mean squared displacement (\enmfluid), confirming that it saturates in the long-time limit~[Fig.~\ref{fig:schematics}(e)]. 

Increasing the target surface area $s_0$ then leads to a transition from caged to diffusive dynamics (Movie~\moviefluid). At, e.g., $s_0 = 6$, cells exhibited ballistic motion at short time scales ($\mathrm{MSD}\sim t^2$, $t\lesssim10^1$), followed by normal diffusion at long time scales ($\mathrm{MSD}\sim t$, $t\gtrsim10^3$) [Fig. \ref{fig:schematics}(e)]. Figure~\ref{fig:schematics}(f) shows the diffusion coefficient $D_s$ (\enmfluid) as a function of $s_0$ and $v_0$ at $\xi=1$. 
Similar to the 2D APE case~\cite{Bi16}, increasing either $v_0$ or $s_0$ sufficiently leads to tissue fluidization and then to an increase of the diffusion coefficient. 

We next ask how the cell aspect ratio influences tissue fluidization. Compared with the cuboidal case ($\xi=1$), both squamous and columnar tissues require a higher $v_0$ to fluidize at the same $s_0$ (Fig.~\siheight~\cite{SI}). Figure~\ref{fig:schematics}(g) shows the boundary between the solid-like and fluid-like regimes as a function of $\xi$ and $s_0$. At fixed $s_0$, increasing $\xi$ drives the tissue from a solid-like to a fluid-like state and, at sufficiently large $\xi$, back to a solid-like state. As $v_0$ increases, the optimal height for fluidisation moves towards  more squamous cells.

This behavior is consistent with the reentrat nature of the rigidity transition reported in a passive VAE model, and can be explained by projecting the 3D monolayer model onto 2D~\cite{Sarkar2026-ql}: Assuming the cells are apicobasally symmetric and have volume $1$, the energy [Eq.~\eqref{eq:energy}] can be rewritten as
\begin{equation}\label{eq:2D}
    w_{\rm proj}=\sum_k\xi^{-2/3}(p_k-p_0)^2\>,
\end{equation}
where $p_k$ is the perimeter of cell $k$ normalized by square root of average area, and
\begin{equation}\label{eq:p0}
    p_0=\xi^{1/3}(s_0-2\xi^{2/3})
\end{equation}
is its effective target value. At the same $s_0$, both squamous and columnar tissues correspond to lower $p_0$ than cuboidal ones, explaining their higher fluidization activity threshold. Plotting the fluidization boundary as a function of $p_0$ rather than $s_0$ removes the reentrant behavior [Fig.~\ref{fig:schematics}(h)] and at low $v_0$, the boundary corresponds to the expected 2D model value of $p_0=3.81$.

Interestingly, Eq.~(\ref{eq:2D}) also shows that the cell aspect ratio $\xi$ directly determines the potential energy scale. Therefore, it also sets the height of energy barriers for topological transitions at a given $p_0$ (i.e., barriers are higher at a lower $\xi$).
This explains why, at a chosen $p_0$, the threshold activity $v_0$ for fluidisation is lower in squamous than in columnar cells [Fig.~\ref{fig:schematics}(h)].

{\textit{Scutoids increase cell motility}}---In the standard 2D vertex model, cell height is omitted. One consequence of this is that T1 transitions are implicitly assumed to happen simultaneously on the apical and basal side. However, cells are three dimensional and can assume a scutoid shape~\cite{GomezGalvez18,Rupprecht2017-lk}, with a different number of apical and basal neighbors. This allows for partial T1 transitions, where a cell (perhaps transiently) exchanges neighbors on one side but not the other [Fig.~\ref{fig:schematics}(d)]. 

To determine if the simplified 2D picture is sufficient to capture tissue dynamics, we compare self-propelled model tissues that allow for scutoids to those that only allow complete T1 transitions. In the latter, a T1 only happens if both the apical and basal edge length are below a threshold $l_{\rm th}=0.01$ at the same time.

Again first focusing on the case of $s_0=6$, $v_0=0.5$, $\xi=1$, we find that MSD in the scutoid model is significantly higher than in the complete-T1 model [Fig.~\ref{fig:scutoid}(a) and Movies~\moviefluid\ and \moviecomplete], suggesting partial topological transitions can significantly increase cell motility. Moreover, the scutoid model also fluidizes at a lower $p_0$ than the complete-T1 model [Fig.~\ref{fig:scutoid}(b)].

As a measure of the relative difference in motility, we compute $\Delta D_s/D_s^{(s)}=\left(D_s^{(s)}-D_s^{(c)}\right)/D_s^{(s)}$, where $D_s^{(s)}$ and $D_s^{(c)}$ are, respectively, the diffusion coefficients of the scutoid and complete-T1 models. Plotting $\Delta D_s/D_s^{(s)}$ as a function of $p_0$ and $v_0$ confirms that the scutoid model is consistently more motile and fluidizes at lower activities than the complete-T1 model [Fig.~\ref{fig:scutoid}(c)]. 

To better understand why allowing scutoid configurations enhances cell motility, we consider a simplified system in which only a single cell self-propels, with a fixed propulsion direction. In the complete-T1 model at, e.g. $t=900$, the self-propelled cell remains in a connected cluster with all of its initial neighbors. Thus, rather than moving relative to its surroundings, the cell advances by dragging its local neighborhood along with it [Fig.~\ref{fig:scutoid}(d) and Movie~\moviesinglecomplete]. By contrast, in the scutoid model, the self-propelled cell has lost several of its initial neighbors from the connected cluster by the same time [Fig.~\ref{fig:scutoid}(d) and Movie~\moviesinglescutoid]. Scutoid-mediated neighbor exchanges therefore allow the cell to move more easily through the tissue, rather than pulling a region along with it.

We quantify this by plotting $n_g(t)$, the fraction of starting neighbors that still form a connected cluster with the self-propelling cell, as a function of time [Fig.~\ref{fig:scutoid}(e); see \enmsingle\ for calculation details]. This fractional cluster size decays more rapidly in the scutoid than in the complete-T1 model [Figs.~\ref{fig:scutoid}(e) and (f)]. Moreover, the cell in the scutoid model moves faster than its complete-T1 counterpart  [Figs.~\ref{fig:scutoid}(g) and (h)].

Altogether, these suggest that the greater freedom allowed by partial T1 transitions and the resulting scutoid shapes enables cells to more easily separate from their neighbors. This separation is necessary for tissue fluidization and diffusive cell motion. Therefore, scutoids allow for greater cell motility in comparison to the complete-T1 case implicitly assumed in the 2D model.


\begin{figure}[tb!]
    \includegraphics[width=\linewidth]{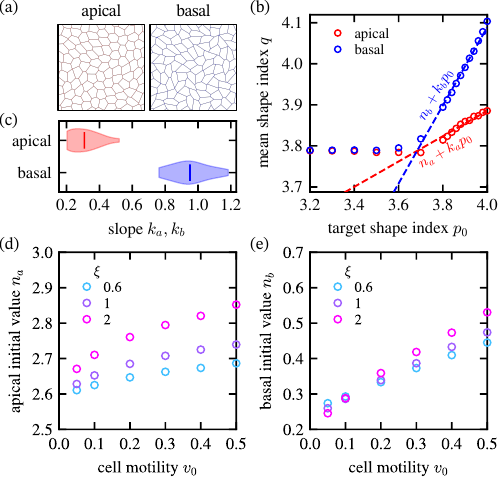}
    \caption{\label{fig:xi} 
    (a) Representative snapshots of the apical and basal side for $p_0=4.0$, $v_0=0.1$, and $\xi=1$. (b) Mean shape index as a function of the target shape index $p_0$ at $\xi=1$ and $v_0=0.1$. The dashed lines denote linear fits to the apical and basal data for simulations with $p_0\geq3.8$. (c) Distribution of the fitted apical and basal slope $k_{l\in\{a,b\}}$, combining data for $\xi\in\{0.6,1.0,2.0\}$ and $v_0\in\{0.05,0.1,0.2,0.3,0.4,0.5\}$. (d),(e) Dependence of the fitted initial value $n_{l\in\{a,b\}}$ on $v_0$ at different $\xi$ for the apical and basal sides, respectively.
    }
\end{figure}
{\textit{Apicobasal asymmetry}}---Besides assuming an apicobasaly symmetric topology, 2D tissue models also implicitly assume a symmetric geometry. However, we find that in the 3D model, cell shapes can differ markedly between the apical and basal side [Fig.~\ref{fig:xi}(a)]. We focus on the average measured shape index (perimeter over root area) on the apical and basal sides. As long as the tissue is solid-like, the apical and basal shape indices match. However, as the target shape index is increased and the tissue fluidizes, their values start to diverge [Figs.~\ref{fig:xi}(b) and \siapicobasal~\cite{SI}]. The basal side, which directly experiences the active forces, exhibits a higher shape index, indicative of a more disordered tiling. By contrast, these forces are transmitted to the apical side indirectly through the cell's passive mechanical response, plausibly accounting for its lower shape index and more ordered organization.

We then fit a simple linear function $q_l=n_l+k_l p_0$ to the measured shape indices vs the target shape index for tissues with $p_0\geq 3.8$. Here, $l\in\{a,b\}$ for the apical and basal side, respectively [Fig.~\ref{fig:xi}(b)]. Scanning over $v_0$ and $\xi$, we find that the slope of the fitted functions does not strongly depend on model parameters, but does differ between the apical and basal sides [Fig.~\ref{fig:xi}(c) and \silinear~\cite{SI}]. Specifically, we obtain $k_a\approx0.31\pm0.09$ and $k_b\approx0.95\pm0.12$, confirming that the apical measured shape index increases more slowly than the basal one with increasing $p_0$. 

Both the apical and basal initial values $n_l$ increase with cell motility $v_0$ [Figs.~\ref{fig:xi}(d) and (e)]. However, on the apical side, the initial value also increases noticeably with the aspect ratio [Fig.~\ref{fig:xi}(d)]: Given the same $p_0$ and $v_0$, a columnar cell has a smaller apical shape index than a squamous one, perhaps indicative of the greater distance between the apical and the basal surface and thus weaker mechanical coupling. On the other hand, the basal initial value is mostly insensitive to the cell height [Fig.~\ref{fig:xi}(e)]. Therefore, although experiments usually image the apical side, it is the basal side that better convey information about motility at a given target perimeter (or vice versa), regardless of cell height.

{\textit{Discussion}}---The dynamics of epithelial monolayers are often studied using simplified two-dimensional models. Here, we investigated how the third dimension affects the dynamics of a monolayer of self-propelled cells and identified several effects that are absent from conventional 2D descriptions. First, cell height plays a key role in tissue fluidization. As cells transition from squamous to cuboidal to columnar shapes, the tissue can exhibit a reentrant solid--fluid--solid transition, with the fluid state favored at intermediate cell heights (Fig.~\ref{fig:schematics}). Second, in 3D, cells can undergo partial topological transitions in which neighbors are exchanged on only the apical or basal side, giving rise to transient scutoid configurations. Allowing these transitions substantially enhances cell motility compared with a model restricted to complete T1 transitions (Fig.~\ref{fig:scutoid}). Third, self-propulsion breaks apicobasal symmetry, with the basal side exhibiting a more disordered organization than the apical side. Moreover, our results suggest that basal cell geometry provides a more direct readout of the mechanical and motility properties of epithelial tissues than the experimentally more accessible apical side (Fig.~\ref{fig:xi}).

\begin{acknowledgments}
{\textit{Acknowledgments}}---We thank Marko Popović, Ana-Sunčana Smith, and all members of the Theoretical biophysics group at JSI for helpful discussions. We acknowledge the financial support from the Slovenian Research and Innovation Agency (research projects J1-60013 and J1-70062, development funding pillar RSF-0106, and research core funding P1-0055).
\end{acknowledgments}

\bibliography{manuscript.bib}

\begin{thebibliography}{38}
\bibitem [{SI()}]{SI}%
  \BibitemOpen
  \href@noop {} {\enquote {\bibinfo {title} {See {S}upplemental {M}aterial at [url will be inserted by publisher] for {F}igures {S}1-{S}3 and {M}ovie captions.}}\ }\BibitemShut {NoStop}%
\end{thebibliography}
\newpage
\clearpage
\begin{center}
\textbf{End Matter}
\end{center}

\noindent
\section*{Appendix A: Model non-dimensionalization}
The tissue potential energy describes elasticity of cell volume and cell surface area:
\begin{equation}
W=\sum_k\left[K_V\left(V_k-V_0\right)^2+K_S\left(S_k-S_0\right)^2\right],
\label{eq:Energy}
\end{equation}
where $V_k$ and $S_k$ are the volume and surface area of cell $k$, respectively ($V_0$ and $S_0$ denote their target values), whereas $K_V$ and $K_S$ are the moduli associated with volume-and-surface-area elasticity. 

We define the dimensionless energy $w=W/\left (K_SV_0^{4/3}\right )$, dimensionless volume $v_k=V_k/V_0$, dimensionless surface area $s_k=S_k/V_0^{2/3}$, dimensionless target surface area $s_0=S_0/V_0^{2/3}$, and dimensionless volume elasticity modulus $k_v=K_VV_0^{2/3}/K_S$.

Vertices follow overdamped dynamics, described by the overdamped equation of motion, which is its full dimensional form reads
\begin{equation}
    \frac{1}{\mu}\dot{\mathbf R}_i=\mathbf F_i^{cons}+\mathbf F_i^{act}+\mathbf F_i^{aux},
\label{eq:vertex_dim_eom}
\end{equation}
where $\mu$ is the vertex mobility, ${\mathbf R}_i$ is the dimensionful position of vertex $i$, the right-hand-side terms represent the three force contributions, and the overdot here indicates the derivative with respect to dimensionful time. We divide Eq.~\eqref{eq:vertex_dim_eom} by $K_SV_0$ to render it dimensionless. Therefore, our dimensionless time is measured in units of $(K_SV_0^{2/3}\mu)^{-1}$.

\section*{Appendix B: Self-propulsion \& auxiliary force}
The second right-hand-side term in Eq.~(\ref{eq:vertex_eom}) represents active forcing due to basal cell self-propulsion. As noted in the main text, each cell $k$ is assigned a polarity angle $\theta_k$, which undergoes angular diffusion, described by  $\partial_t\theta_k=\eta_k(t)$, with noise statistics $\left\langle \eta_k(t)\right\rangle =0$ and $\left\langle \eta_k(t)\eta_l(t')\right\rangle=2D_r\,\delta_{kl}\,\delta(t-t')$, where $D_r$ is the rotational diffusion coefficient. Since the equations of motion are written for vertices rather than cell centers, we compute direction of active force on each basal vertex $i$ using the approach of Ref.~\cite{sussman2017cellgpu}: The vertex force is calculated from the average polarity of the cells sharing that vertex as $\mathbf f^{act}_i=\left (1/|\mathcal C(i)|\right )\sum_{k\in\mathcal C(i)}v_0\hat{\mathbf n}_k$, where $v_0$ sets the magnitude of the self-propulsion force, $\hat{\mathbf n}_k=(\cos\theta_k, \sin\theta_k)$ is the polarity vector of cell $k$, and $\mathcal C(i)$ is the set of cells surrounding the vertex $i$.

The third, auxiliary term in Eq.~\ref{eq:vertex_eom} has the form $\mathbf f_i^{aux}=-\sum_k^{N_c} \sum_{l\in \{a,b\} } 2 k_A \mathrm H(A_0-A^l_k)(A^l_k-A_0)\nabla_iA^l_k$, where $\mathrm H$ denotes the Heaviside step function, $A_k^l$ is the area of the apical ($l=a$) or basal ($l=b$) surface of cell $k$, and $A_0$ is a threshold area below which further area reduction is penalized. We set the corresponding modulus to $k_A=1$ and $A_0=L^2/N_c=\xi^{2/3}$.

\section*{Appendix C: Quantifying fluidization}
The mean squared displacement of the cells is calculated as
\begin{equation}
    \mathrm{MSD}(t)=\frac{1}{\left<A\right>}\left\langle \left| \tilde{\mathbf{r}}_k(t)-\tilde{\mathbf{r}}_k(0) - \Delta \tilde{\mathbf{r}}_{tissue}(t)\right|^2 \right\rangle,
\end{equation}
where $\tilde{\mathbf{r}}_k=(1/|\mathcal V(k)|)\sum_{i\in\mathcal V(k)}\mathbf r_i$ is the centroid of the basal polygon of cell $k$, $\mathcal V(k)$ denotes the vertices of the basal polygon,  $\Delta\tilde{\mathbf{r}}_{tissue}(t)=\langle\tilde{\mathbf{r}}_k(t)-\tilde{\mathbf{r}}_k(0)\rangle$ is the displacement of the tissue's centroid, and the average is taken over cells. For ease of comparison of cell motility between different aspect ratios, we then normalize MSD by the average apical (or basal) cell area $\left<A\right>=L^2/N_c=\xi^{2/3}$. 

To reach the steady state, each simulation is first evolved for $t=t_0=10^4$. We then reset the time origin, $t = 0$, and collect data up to $t=10^5$. The diffusion coefficient is given by $D_s=\lim_{t\to\infty}\mathrm{MSD}(t)/(4t)$. We classify tissues as solid-like if $D_s<2.5\times10^{-5}$ (corresponding to final $\mathrm{MSD}<10$) and as fluid-like above that threshold.

\section*{Appendix D: Single self-propelled cell model}
To acquire the steady state, each simulation is first evolved for $t=t_0=10^4$ using the complete-T1 model. We then reset the time origin, $t = 0$, and collect data up to $t=10^4$. To calculate the size of the connected cluster of active cell $k$'s starting neighbors, let $\mathcal{C}_k(0)$ denote the set of cells sharing an edge with it at $t=0$. At each time $t$, we construct the connected cluster $\mathcal{C}'_k(t)$ those cells from $\mathcal{C}_k(0)$ that remain connected to $k$ through the topological nearest-neighbor graph. We quantify the persistence of the initial neighborhood by the normalized cluster size $n_g(t)=\left|\mathcal{C}'_k(t)\right|/\left|\mathcal{C}_k(0)\right|$. Thus, $n_g=1$ as long as the initial neighborhood remains fully connected, while $n_g<1$ signals the loss of connectivity with at least one initial neighbor. The uncaging time $\tau_g$ in Fig.~\ref{fig:scutoid}(f) is defined as the time at which $n_g(t)$ first falls below 1. Figure~\ref{fig:scutoid}(g) shows the average of the displacement $\Delta {\mathrm{r}}(t)=|\tilde{\mathrm{r}}_k(t)-\tilde{\mathrm{r}}_k(0)|$ between different simulations, where $k$ is now the index of the self-propelling cell. Lastly, Fig.~\ref{fig:scutoid}(h) shows the distribution of self-propelled cell velocities obtained from different simulations by fitting $\Delta \mathrm{r}(t)=v \times t$. 

\end{document}


\newcommand{\zc}{\zeta_{\rm{c}}}
\newcommand{\ze}{\zeta_{\rm{e}}}

\newcommand{\figdroplet}{3}

\preprint{}

\title{
\textbf{Supplemental Material: \\ {Full Three-dimensional Cell Geometry Matters for Epithelial Fluidization}}}
\author{Toshikaze Chiba}
\affiliation{Jožef Stefan Institute, Jamova 39, SI-1000 Ljubljana, Slovenia.}

\author{Jan Rozman}
\affiliation{Jožef Stefan Institute, Jamova 39, SI-1000 Ljubljana, Slovenia.}
\affiliation{University of Ljubljana, Faculty of Mathematics and Physics, Jadranska 19, SI-1000 Ljubljana, Slovenia.}

\author{Matej Krajnc}
\affiliation{Jožef Stefan Institute, Jamova 39, SI-1000 Ljubljana, Slovenia.}

\date{\today}

{
\let\clearpage\relax
\maketitle
}

\section*{Supplementary figures}

\begin{figure*}[h]
    \centering
    \includegraphics[width=\textwidth]{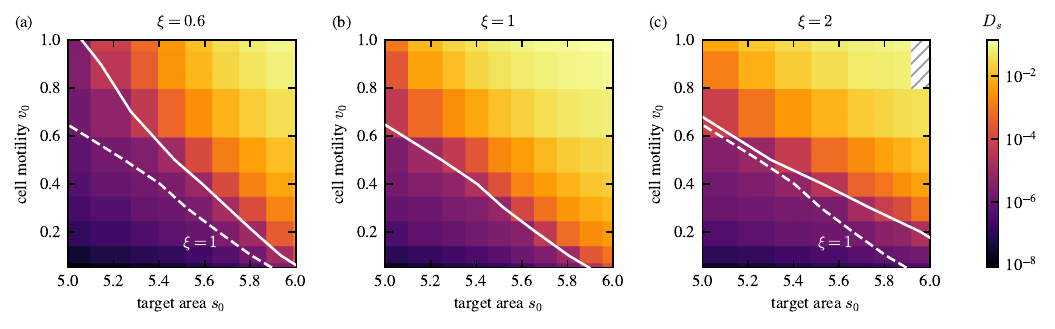}
    \caption{Heatmaps of diffusion coefficient at different $\xi$ for $\xi=0.6$ (a), $\xi=1$ (b), and $\xi=2$ (c). The bold lines indicate the solid–fluid boundary. For ease of comparison, the dashed lines on (a) and (c) show the solid-fluid boundary at $\xi=1$. The hatched region indicates where the simulations are numerically unstable. 
    }
\end{figure*}

\begin{figure*}[h]
    \centering
    \includegraphics[width=\textwidth]{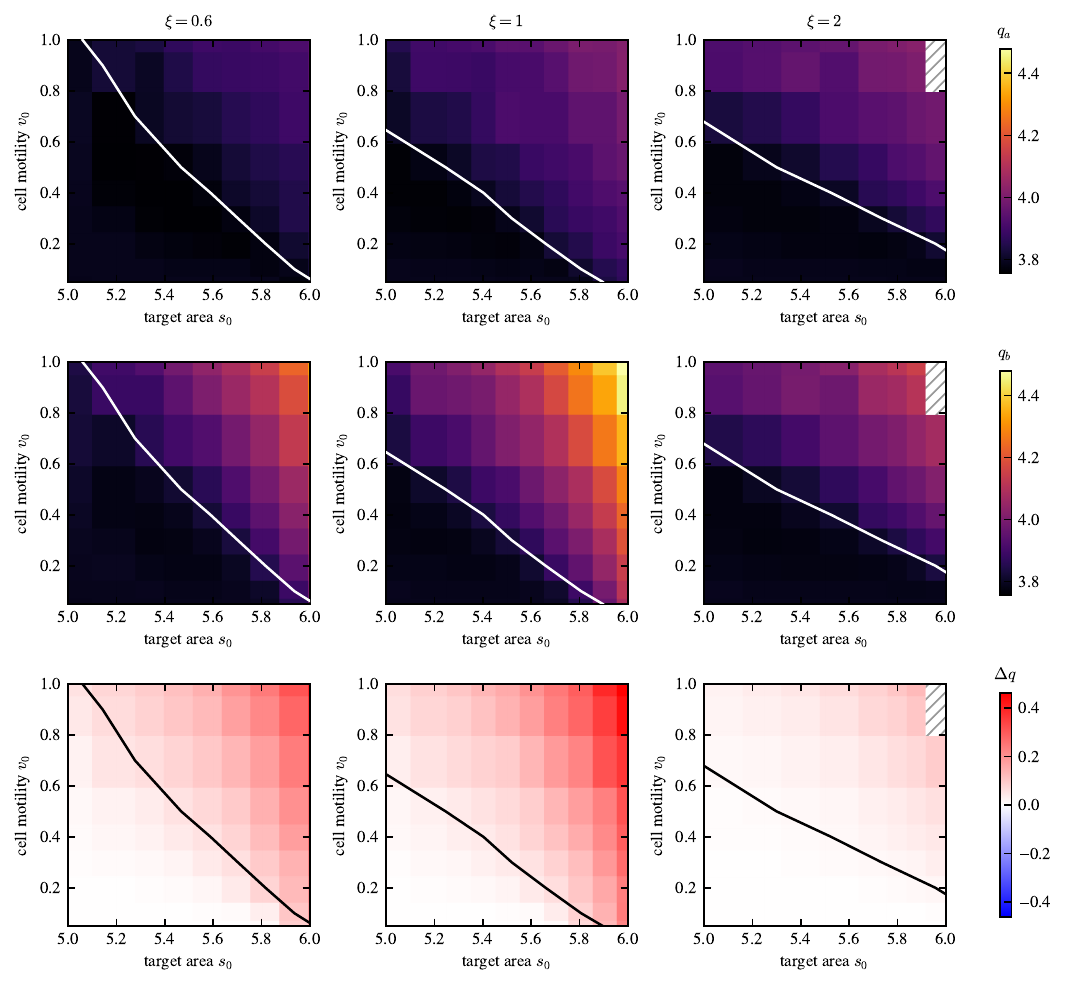}
    \caption{Heatmap of the mean measured shape index: $q_a$ and $q_b$ denote the apical and basal mean shape indices, respectively, and $\Delta q = q_b - q_a$ denotes the difference in shape index between the basal and apical sides. The bold line indicates the solid–fluid boundary. The hatched region indicates where the simulations are numerically unstable.
    }
\end{figure*}

\begin{figure*}[h]
    \centering
    \includegraphics[width=0.6\textwidth]{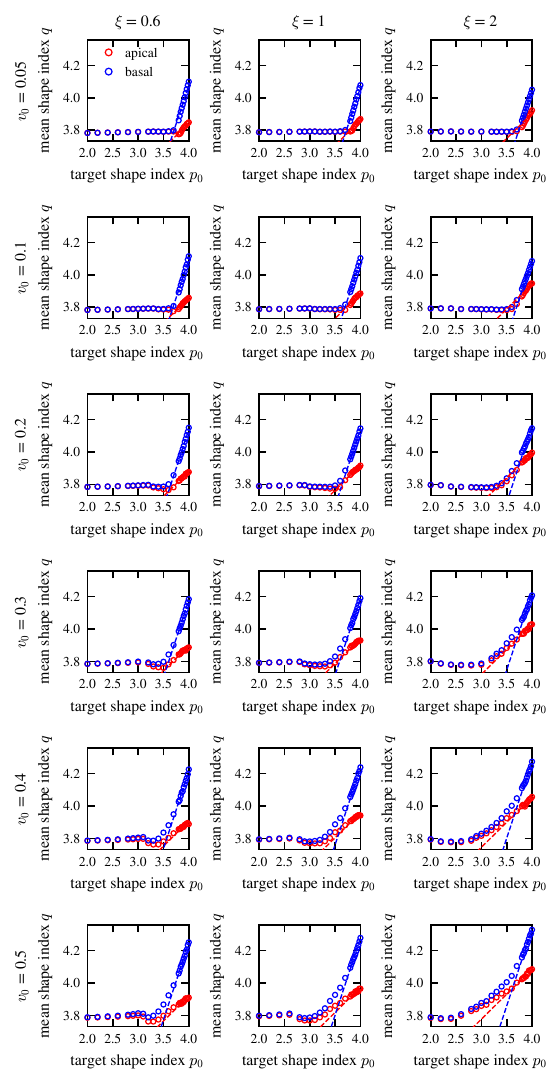}
    \caption{The mean measured shape index as a function of the target shape index for different parameter values. 
    }
\end{figure*}

\clearpage
\newpage

\section*{Movie captions}
	\noindent Movie S1. Model tissue evolution for $v_0=0.5$, $\xi=1$, and $s_0=5$, showing a solid-like system. The starting neighbors of one cell (orange) are colored yellow. Note that all movies are centered on the tissue's centroid ($\Delta\tilde{\mathbf{r}}_{tissue}(t)$ in Appendix C).\\ 
    
	\noindent Movie S2. Model tissue evolution for $v_0=0.5$, $\xi=1$, and $s_0=6$, showing a fluid-like system. The starting neighbors of one cell (orange) are colored yellow.\\ 
    
	\noindent Movie S3. Model tissue evolution for $v_0=0.5$, $\xi=1$, and $s_0=6$, using the complete-T1 topological transition model. The starting neighbors of one cell (green) are colored yellow.\\ 
    
	\noindent Movie S4. Model tissue evolution with a single self-propelling cell (green), using the complete-T1 model. The self-propelling cell's starting neighbors are colored yellow.\\ 
    
	\noindent Movie S5. Model tissue evolution with a single self-propelling cell (orange), using the scutoid model. The self-propelling cell's starting neighbors are colored yellow.\\